\documentclass[
    aps,
    prl,
    reprint,
    showkeys,
    amsmath,
    amssymb,
    superscriptaddress,
]{revtex4-2}

\usepackage{graphicx}
\usepackage{hyperref}
\usepackage{microtype}

\begin{document}

\title{Distributional signatures of computation: inferring mechanisms from computation time distributions}

\author{C\'eline van Valkenhoef}
\affiliation{University of Vienna, Faculty of Physics, Boltzmanngasse 5, 1090 Vienna, Austria}

\author{Catherine Schuman}
\affiliation{Department of Electrical Engineering and Computer Science,
University of Tennessee, Knoxville, Tennessee, USA}

\author{Nikolaus Kriegeskorte}
\affiliation{Departments of Psychology, Neuroscience and Electrical Engineering, Zuckerman Mind Brain Behavior Institute,
Columbia University, New York, NY 10027, USA}

\author{Philip Walther}
\affiliation{University of Vienna, Faculty of Physics, Boltzmanngasse 5, 1090 Vienna, Austria}
\affiliation{University of Vienna, Research Network for Quantum Aspects of Space Time (TURIS), Boltzmanngasse 5, 1090 Vienna, Austria}

\collaboration{\normalfont\textit{Correspondence:}
\href{mailto:celine.vanvalkenhoef@univie.ac.at}
{celine.vanvalkenhoef@univie.ac.at};
\href{mailto:philip.walther@univie.ac.at}
{philip.walther@univie.ac.at}}

\begin{abstract}
Computational complexity theory provides a theoretical framework for analysing abstract models of computation, whereas computational systems operate as finite physical implementations whose behaviour is shaped by their architecture and physical substrate. In many such systems, computational organisation cannot be directly characterised from algorithms or internal representations. Building on the use of reaction time distributions to constrain mechanisms in cognitive science, we introduce an empirical framework for investigating computation in finite physical systems by inferring computational mechanisms from observable computation time distributions. The framework combines empirical benchmarking, which characterises computational efficiency through the scaling of computation time with problem size, with distributional analysis, which treats computation time distributions as signatures of the stochastic dynamics generated by different computational mechanisms. We demonstrate the framework using a scalable visual search paradigm, comparing human performance with three computational models that employ distinct mechanisms of evidence accumulation. Benchmarking reveals broadly similar scaling behaviour for the models and human observers, whereas distributional analysis distinguishes between these models. More generally, we show that computation time distributions provide an empirical level of description linking observable computational behaviour to underlying computational mechanisms. This framework complements computational complexity theory by extending the empirical study of computation from measuring computational performance to characterising computational organisation. It provides an empirical approach for comparing finite computational systems across diverse physical implementations, including biological, classical, neuromorphic, and quantum systems, without requiring access to their internal algorithms or architectures. 
\end{abstract}

\keywords{distributional signature,
computational mechanisms,
computation time distributions,
physical computation,
finite-scale computation}

\maketitle

Computational complexity theory has been successfully used to classify problems according to the resources required by abstract computational models as problem size approaches infinity. By analysing the asymptotic resource requirements of algorithms, it provides a theoretical measure of computational efficiency. \cite{arora2009computational} However, the emergence of computational paradigms such as quantum and neuromorphic computing has broadened the range of computational architectures available. These architectures are often associated with different computational models and complexity classes, making it increasingly difficult to compare their practical computational performance. \cite{indiveri2011neuromorphic, schuman2022opportunities, nielsen2010quantum} In addition, other computational systems, such as human cognition and modern AI systems, operate as finite physical implementations whose computational behaviour is shaped by the opportunities and constraints arising from their architecture and substrate. Many of these systems are also adaptive, such that their computational organisation cannot be readily characterised using existing approaches based on explicit algorithms and abstract computational models. \cite{kandel2000principles, maclennan2004natural, arora2009computational}

These developments motivate two central questions. How can we compare computational systems whose internal organisation is complicated or inaccessible? And how can finite physical implementations of computation be compared empirically?

In this work, we build on the use of reaction time distributions to constrain mechanisms in cognitive science. \cite{wolfe2010reaction, fific2008studying} We extend this principle by developing an empirical framework for investigating the organisation of computation in finite physical systems from observable computational behaviour. Central to this framework is the concept of computational mechanisms, which provide an intermediate level of description connecting the physical resources available for computation to its observable behaviour. Computational mechanisms thus describe how computation is organised within finite physical systems.

The framework consists of several complementary components. Benchmarking characterises how computation time scales with problem size. Distributional analysis extends this by treating computation time distributions as observable signatures of stochastic computational mechanisms. These mechanisms arise from the physical implementation of computation, generating characteristic stochastic dynamics and distributional signatures. The reverse direction constitutes empirical inference, whereby observed distributions are used to infer the underlying computational mechanisms.

We demonstrate the framework using visual search, comparing human performance with computational models. More generally, the framework provides an empirical approach for investigating computational mechanisms in finite physical systems, independent of their underlying architecture or substrate.

\section*{From abstract to physical computation}

Alan Turing introduced a model of computation by analysing the work of human ‘computers' and reducing their methods to elementary steps. This led to the formulation of what later became known as the Turing machine, a canonical model of computation. The Turing machine provides an algorithmic model of computation that abstracts away from architectural constraints and physical limitations. It therefore defines an abstract notion of computation, in which algorithms are analysed independently of the physical systems used to implement them. The running time of a computation is commonly measured by the number of elementary operations performed by the machine before reaching the halting state. Running time is typically analysed asymptotically as a function of input size \textit{n}, providing a basis for comparing the complexity of computational problems. \cite{turing1936computable, moore2011nature, arora2009computational}

Whilst the Turing machine captures the abstract notion of algorithmic computation, it does not explicitly represent the architectural or physical constraints of real computational systems. In particular, it assumes deterministic and discrete state transitions, noiseless operation, and effectively unlimited memory. Real computational systems, however, may differ substantially from these assumptions, and their performance is shaped by finite resources and physical constraints. \cite{maclennan2007review, landauer1961irreversibility, bennett1982thermodynamics}

In this work, we are primarily concerned with how computation is organised in finite physical systems. Algorithms determine what sequence of operations is performed, architectures determine how those operations are organised, and substrates determine the physical processes through which they are realised. Together, these define the physical implementation of a computational system, whose properties determine the resources available and, consequently, the computational behaviour that can be achieved in practice. Computational mechanisms are not determined by the algorithm or physical implementation alone, but emerge from how an algorithm exploits the resources, opportunities, and constraints afforded by a particular physical system. \cite{marr2010vision}

Physical computation takes different forms depending on how information is represented and processed, and on the physical principles governing computational evolution. Digital and analogue computation differ in how information is represented and manipulated, whereas classical and quantum computation differ in the physical laws governing computational evolution. Mixed-signal computation combines digital and analogue processing to exploit the complementary advantages of both approaches. \cite{shannon1938symbolic, lloyd2000ultimate, indiveri2011neuromorphic, kandel2000principles, nielsen2010quantum}

Understanding physical computation therefore requires considering not only abstract algorithms and models of computation, but also the physical processes through which they are realised. This perspective motivates the empirical framework developed below, which investigates computation in finite physical systems through their observable computational behaviour.

\section*{Benchmarking finite computational systems}

The limitations of asymptotic analysis motivate an alternative approach based on empirical benchmarking of finite computational systems. \cite{mcgeoch2012guide,ahuja1996use, jain1990art} This approach characterises computational performance by measuring computation time across a range of problem sizes. From these measurements, we estimate the scaling relationship between computation time and problem size, which provides an empirical measure of computational efficiency. To extract the scaling behaviour, we fit scaling functions to measurements of computation time as a function of problem size, such as constant, logarithmic, linear, polynomial, or exponential. Thus, empirical benchmarking captures performance in a finite-size regime that is realistically achievable by the computational system. Crucially, it can be applied across diverse computational paradigms without requiring knowledge of the underlying implementation, enabling direct comparison of systems ranging from human cognition and artificial neural networks to analogue and quantum computational approaches.

To illustrate this, consider benchmarking human cognition. In this case, computation time can be approximated by reaction time (RT), a standard measure in cognitive science defined as the time between stimulus and response. \cite{donders1868over, sternberg1969discovery, luce1991response} Reaction time therefore serves as an operational measure of computation time. To apply this approach, tasks must be defined such that they can be solved by human participants over a range of problem sizes, allowing the collection of sufficiently large datasets for scaling analysis.

In cognitive science, visual search provides a well-established paradigm for studying the computational dynamics underlying information processing and evidence accumulation. In a typical task, observers search for a target amongst distractors whilst RT is measured. \cite{wolfe2017five,chan2013visual,eckstein2011visual} This paradigm, typically focusing on modest numbers of distractors, has led to many important insights and theories. \cite{treisman1980feature,thorpe1996speed,wolfe2021guided, duncan1989visual, lleras2020target} Classical visual search analyses primarily characterise performance using mean RT as a function of set size, thereby compacting noisy datasets into a clear relationship between mean RT and problem size. However, this compression discards distributional information that contains mechanistic insight into the underlying computation. \cite{palmer2011shapes, heathcote1991analysis,rousselet2018reaction} In addition, the use of RT measures aggregated across participants may distort the shape of RT distributions. Finally, traditional paradigms are difficult to scale to large input sizes, with set sizes often limited to a few tens of items and corresponding RTs of only a few seconds. These limitations motivate a visual search paradigm suitable for combining finite-scale benchmarking with distributional analysis.

To demonstrate the framework, we introduce a novel visual search paradigm designed to support scalable finite-scale benchmarking and distributional analysis of computation time. Our target is an emoticon constructed from $7\times7$ black and white pixels, which itself sits within a background of black and white pixels. Thus, instead of placing the target and distractors in an empty space, we define the input size as the total number of pixels. We vary the input size from 400 to nearly 320,000 image locations and measure RTs across this range. 

We compare human performance with three representative computational models spanning different computational mechanisms. These include two versions of the drift diffusion model (DDM), which accumulate information through sequential sampling, whereby evidence is integrated over time until a decision threshold is reached. \cite{ratcliff1978theory, corbett2020diffusion, ratcliff2016diffusion} The simplest version, the random DDM (rDDM), draws sample locations uniformly and independently at random without using memory to guide sampling (i.e. without memory of previously visited locations). Although it maintains a global posterior map and thus possesses spatiotemporal memory, it does not use this information to guide sampling. To make search more efficient, the guided DDM (gDDM) samples at the location currently most likely to contain the target. Because the gDDM fully exploits the posterior to guide sampling, it represents a fully guided spatiotemporal memory model. 

In both models, sampling and updating continue until the posterior exceeds a decision threshold at any location, at which point the model reports detection at that location. RT is defined as the total number of samples required to reach the decision threshold. Whilst these cognitive models capture essential aspects of evidence accumulation, they do not operate directly on the input image and therefore do not provide an image-computable mechanistic account. We therefore also consider a mechanistic class of models: artificial neural networks.

\begin{figure*}[t]
\centering
\includegraphics[width=11.4 cm, keepaspectratio]{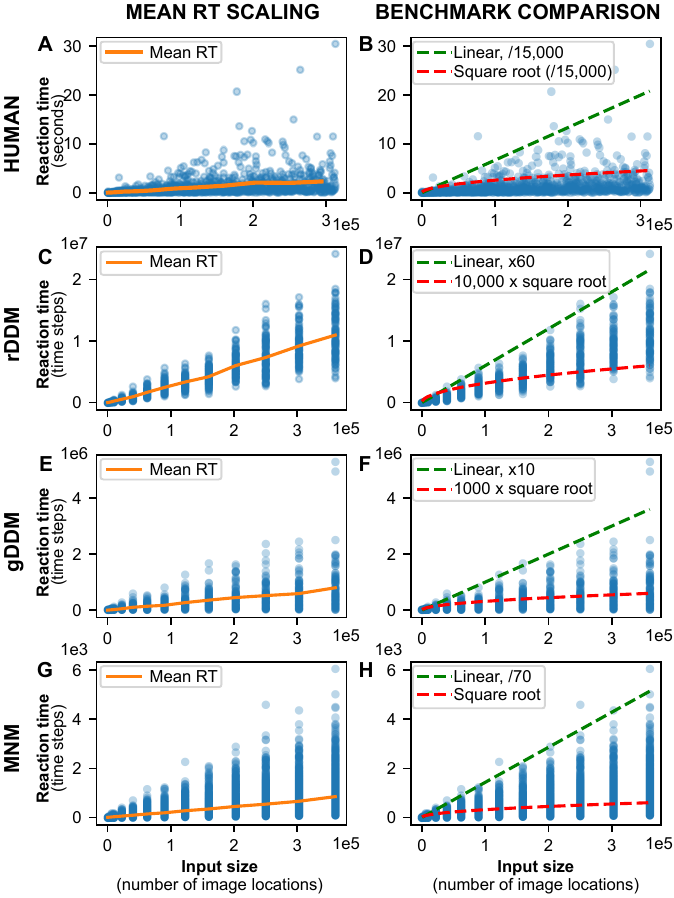}
\caption{\textbf{Mean RT scaling and benchmarking comparisons for human data and computational models. }The left column shows the RT distributions and mean trends for the human participant and the three models, each exhibiting approximately linear increases in mean RT with input size. The right column shows linear and square-root benchmark curves superimposed on the distributions. Arbitrary scaling factors, indicated in the legends, are required to align the benchmark curves with the data. Thus, although benchmarking constrains broad scaling behaviour, interpretation depends strongly on scaling assumptions and does not uniquely determine the underlying computational mechanism.} 
\label{fig1}
\end{figure*}

As our third model, we introduce a simple mechanistic model based on a shallow spiking neural network architecture (mechanistic neural network model, MNM). \cite{maass1997networks, plank2022case} Visual search is performed by repeatedly placing a foveated field of view over the image, allowing multiple locations to be processed simultaneously. Rather than maintaining a global posterior, the model accumulates evidence locally through repeated evaluation and confirmation of locations. Initial placement of the field of view is random, while confirmation, if triggered, occurs at the same location, thereby providing local memory. If the confirmation step is positive, verification occurs, in which the detected coordinates are compared to the true target location. If they match, the search terminates; otherwise, the model returns to exploration. RT is defined as the total number of evaluation and confirmation steps required to achieve successful verification, together with a verification cost that scales with image size.

We refer to this model as mechanistic because it implements visual search using explicitly defined and interpretable computational components within a parallel, noisy architecture. Unlike drift diffusion models, which operate on abstract evidence samples, the MNM processes visual input directly and is therefore image-computable. The MNM thus provides a simple and interpretable framework that bridges elements of cognitive and neural network models. \cite{kriegeskorte2015deep} Detailed descriptions of the search paradigm and the models are provided in the Methods and Supplementary Information.

Mean-based RT analysis forms the classical approach to visual search, where the slope of mean RT as a function of set size is interpreted as a measure of search efficiency. \cite{treisman1980feature,wolfe2021guided} More generally, benchmarking approaches compare scaling behaviour of computational systems against known complexity trends. \cite{mcgeoch2012guide,ahuja1996use} Within the framework, such analyses constrain broad computational efficiency by characterising how computation time scales with input size.

To compare the models to each other and to the human data, we first carry out a finite-scale mean-based and benchmarking analysis. The left column of Fig. \ref{fig1} shows the RT distributions together with their mean trends marked in orange for a large dataset collected from a single participant together with the datasets from the rDDM, gDDM, and MNM. For all systems, the mean RT increases approximately linearly with input size. Although the slopes differ across systems, these differences are difficult to interpret because model time steps are not intrinsically comparable either to one another or to human RTs measured in seconds.

In the right column of Fig. \ref{fig1}, linear and square-root benchmark curves are superimposed on the distributions. The linear benchmarks were scaled to approximate the upper bounds of the distributions, with the exact scaling factors indicated in the legends. For the human data (Fig. \ref{fig1}B), applying the same scaling factor to both benchmarks suggests that the linear and square-root curves respectively approximate the upper envelope and a region of high density within the distribution. However, for the DDMs (Fig. \ref{fig1}D,F), substantially larger scaling factors are required for the square-root benchmarks than for the linear benchmarks. In contrast, for the MNM  (Fig. \ref{fig1}H) the scaling factor required for the square-root benchmark is smaller than that for the linear benchmark.

Overall, we conclude that for this task, both mean RT and upper-bound behaviour show broadly linear increases with input size across the human data and models. However, because the units of model time are not directly comparable to one another or to human RTs measured in seconds, we cannot directly relate differences in slope to underlying computational mechanisms. More generally, although benchmark curves may appear meaningful, their interpretation depends strongly on arbitrary scaling choices. Finite-scale benchmarking therefore constrains broad computational efficiency and scaling behaviour, but does not uniquely distinguish between different underlying computational mechanisms. To resolve these ambiguities, we now extend the analysis to the full RT distributions and their profiles.

\section*{Distributional signatures of computation}

Empirical benchmarking typically involves repeated measurements of computation time at each problem size. In many computational systems, repeated measurements produce a range of computation times rather than a single value owing to stochastic variability. These computation time distributions reflect the stochastic dynamics of the computation, including how information is sampled, accumulated, and transformed. Because these dynamics arise from the underlying computational mechanism, computation time distributions provide information about computational mechanisms that is not captured by mean behaviour alone. \cite{gardiner2009stochastic, townsend1983stochastic, van1983stochastic}

\begin{figure*}[t]
\centering
\includegraphics[width=11.4 cm, keepaspectratio]{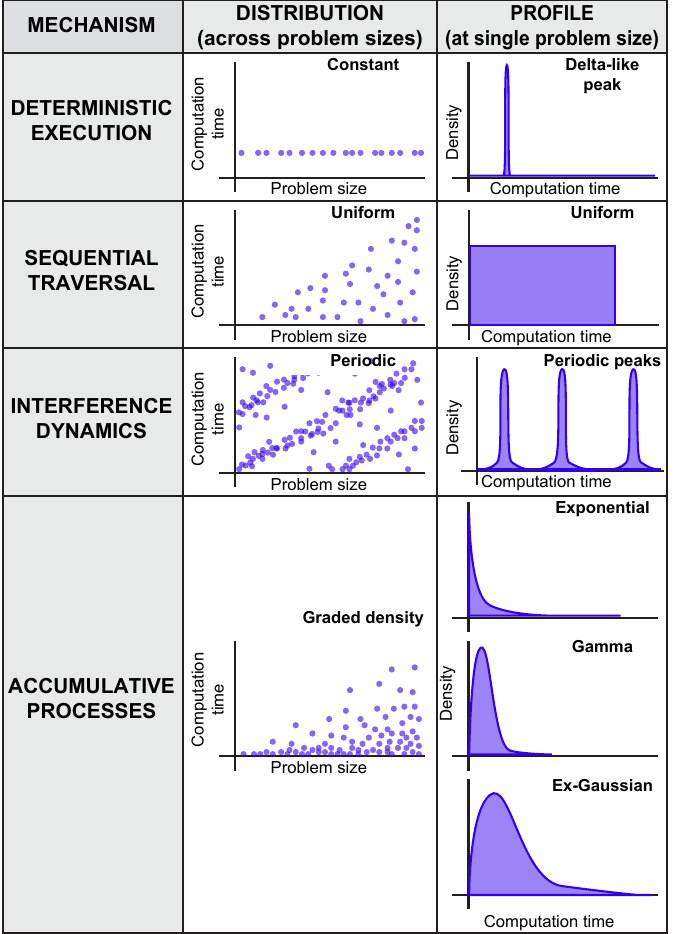}
\caption{\textbf{Schematic illustration of computation time distributions and profiles across computational mechanisms.} The middle column shows variability in computation time across problem sizes, whilst the right column shows the corresponding profile at a fixed problem size. \textbf{Row 1:} An illustrative deterministic mechanism produces fixed computation times, resulting in a constant distribution and a delta-like profile. \textbf{Row 2:} Sequential traversal processes proceed through the problem space in an ordered sequence of computational steps. For uniformly distributed solution locations, this produces uniformly distributed stopping times. \textbf{Row 3:} Interference dynamics generate oscillatory computation-time structure and profiles with periodic peaks. \textbf{Row 4:} Accumulative processes produce graded variability structures and positively skewed profiles whose width and skewness increase with the degree of accumulation.}
\label{fig2} 
\end{figure*}

In Fig. \ref{fig2}, we illustrate how representative computational mechanisms give rise to characteristic computation time distributions and their associated profiles. The computation time distributions show the variability in computation time due to the stochastic computational dynamics of the system across problem sizes. Unlike mean-based summaries, the corresponding computation time profiles preserve this variability by considering a cross-section of the distribution at a fixed problem size. In the first row, we consider deterministic execution. As an illustrative example, we consider a deterministic mechanism whose computation time is independent of problem size. This produces a single computation time at every problem size, giving rise to the constant computation time distribution and delta-like profile shown in the figure. The second row shows sequential traversal, with which we mean a mechanism that proceeds through the problem space in an ordered sequence of computational steps, such that the stopping point depends on where the solution is encountered within the traversal. If we assume that the solution location is placed at random within the deterministic path structure, then the stopping times will be distributed uniformly across problem sizes. The related profile also is uniform. Interference dynamics, shown in the third row, exploit the constructive and destructive interactions in wave-like dynamics, resulting in a banded computation time distribution. The resulting profile then shows periodic peaks. 

Finally, accumulative processes, as depicted in row 4, rely on progressive integration of evidence. This results in a graded distribution, where we have a higher density of data at lower computation times across problem sizes. The degree of this gradation depends on the structure and extent of accumulation inherent in the process, which is often related to the extent to which information from previous computational steps is retained. Illustratively, increasing accumulation produces profiles that transition from approximately exponential, through gamma, towards ex-Gaussian form. Increasing accumulation therefore changes the shape of the profile by broadening the distribution and increasing its positive skew. Thus, we see that the different mechanisms have distinct profile shapes. In contrast with upper bounds, which reflect the efficiency of a computation, computation time distributions and profile shapes characterise the underlying mechanism of the computation.

\begin{figure*}[t]
\centering
\includegraphics[width=11.4 cm, keepaspectratio]{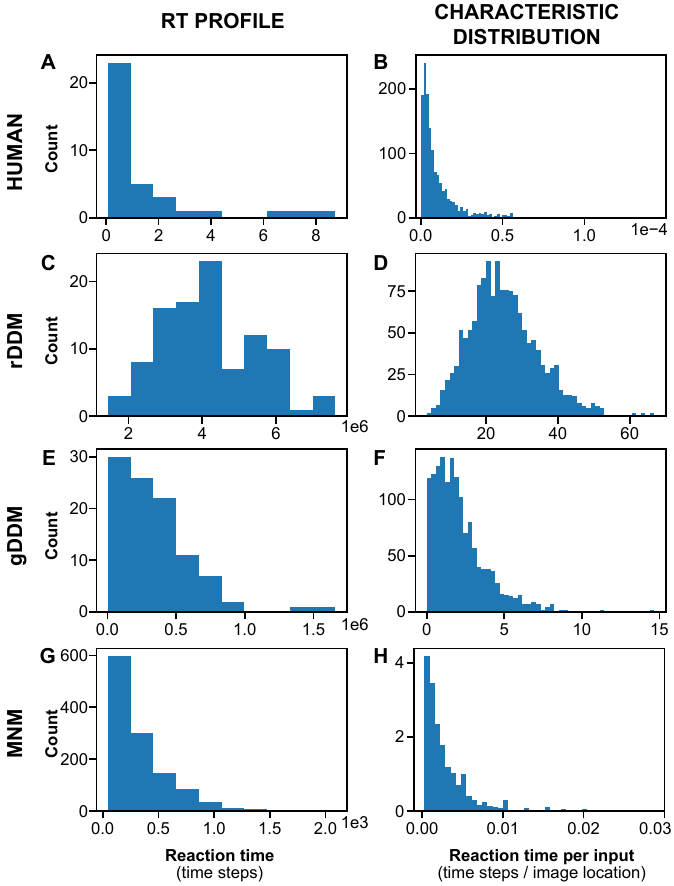}
\caption[RT profiles and characteristic distributions for data and models.] { \textbf{RT profiles and characteristic distributions for data and models. A–B:} Human data from one representative participant, showing an RT profile for 155,000–165,000 image locations, and the characteristic distribution (an estimate of RT profile shape aggregated across input sizes). The characteristic distribution is obtained by dividing each RT by its corresponding input size, allowing comparison of distributional shape across input sizes by expressing time per image location. This normalization assumes approximately linear scaling of reaction time with input size. \textbf{C-D}, \textbf{E-F}, and \textbf{G-H} show the RT profile at 160,000 image locations and the characteristic distribution for the random drift diffusion model (rDDM), guided drift diffusion model (gDDM), and mechanistic neural network model (MNM), respectively. For visual clarity, the long tails of the characteristic distributions in the human data and MNM are truncated along the horizontal axis.}
\label{fig3}
\end{figure*}

Thus, we can extract key features of the distributional structure by considering its profile. However, reliably estimating RT profiles at individual input sizes requires a large amount of data. To address this, we construct a characteristic distribution by divisively normalising RTs by input size, thereby aggregating data across input sizes and removing the input dimension under the assumption of approximately linear RT scaling, as established in Fig. \ref{fig1}. The resulting characteristic distribution is obtained by expressing RT per image location, and can be interpreted as an aggregate estimate of the scale-normalized profile across input sizes. This normalisation allows us to focus on distributional shape, and evaluate how accurately different models capture the human distribution.

We now apply this procedure to the human data and the three computational models. We recall from Fig. \ref{fig1} that the human data and the three models all capture the approximate linear scaling of mean RT with input size, thus confirming that mean-based analyses alone do not distinguish between them. Fig. \ref{fig3} shows the RT profiles and characteristic distributions from the four datasets. The RT profiles in the left column are reflected in their associated characteristic distributions. The characteristic distribution of the human dataset (Fig. \ref{fig3}B) exhibits a high density of observations at short RTs per image location, together with an extended tail of longer RTs. When compared to the models, the MNM (Fig. \ref{fig3}H) provides the closest qualitative match, capturing both the high density of short RTs and the heavy-tailed structure observed in the human data. By contrast, the rDDM (Fig. \ref{fig3}D) shows the greatest deviation, with a shifted distribution and an increasing lower bound that is not observed in human performance. The gDDM (Fig. \ref{fig3}F) captures some aspects of the skew but fails to reproduce the full distributional shape. Together, this qualitative comparison indicates that the MNM reproduces the human distributional shape more closely than the two DDM variants.

For a robust comparison, we want to assess how the models perform comparatively to each other, and thus which model provides the closer match to the observed RT distribution. We compared the models using two complementary approaches that target different aspects of distributional fit. 

After aligning model and human time scales, the Kolmogorov–Smirnov distance assessed residual differences in distributional shape, whereas leave-one-out marginal log-likelihood assessed overall probabilistic agreement while integrating over uncertainty in temporal scaling. Both analyses yielded the same ordering: the rDDM provided the weakest match, the gDDM improved substantially, and the MNM provided the closest match to human RT distributions. Full procedures and statistical results are reported in Methods and the Supplementary Information.

Fig. \ref{fig2} illustrated that accumulative computational mechanisms produce graded computation time distributions, as in human visual search, with associated skewed profiles whose width and skewness depend on how evidence is accumulated over time. The three models considered here employ distinct mechanisms of evidence accumulation, which are reflected in the empirical distributions. The MNM employs dwell-based accumulation of evidence over time, without the need for spatial memory. As such, it generates an intermediate, gamma-like RT profile characteristic of limited-memory accumulative mechanisms. The gDDM maintains a global spatiotemporal evidence map, which it uses for accumulating evidence. This evidence is used both to guide sampling towards the most probable target location and to determine when the decision threshold has been reached. By retaining and exploiting accumulated evidence globally across the search space, the gDDM produces broader and more positively skewed RT profiles than the MNM. The rDDM also maintains a global spatiotemporal evidence map, but unlike the gDDM it does not use this information to guide sampling locations. Sampling therefore remains effectively random, causing the model to revisit locations with low posterior probability and requiring substantially more samples before the decision threshold is reached. This inefficient sampling prolongs evidence accumulation and increases variability in the number of samples required, giving rise to broader and more strongly skewed RT profiles than those observed in the gDDM. Indeed, we see that the differences in how our three models accumulate and exploit evidence during search are reflected in the empirical distributions.

Taken together, these findings demonstrate that computation time distributions provide information about computational mechanisms that is not available from benchmarking alone. While benchmarking constrains computational efficiency, distributional analysis distinguishes between alternative computational mechanisms that exhibit similar scaling behaviour. For the models considered here, the results indicate that a limited-memory, dwell-based, image-computable mechanism provides the closest account of human visual search behaviour observed in our paradigm. More generally, computation time distributions are not merely noise around a mean, but reflect the stochastic dynamics of the computational system. They therefore provide an empirical basis for inferring computational mechanisms from observable behaviour.

\section*{An empirical framework for inferring computational mechanisms}

\begin{figure}
\centering
\includegraphics{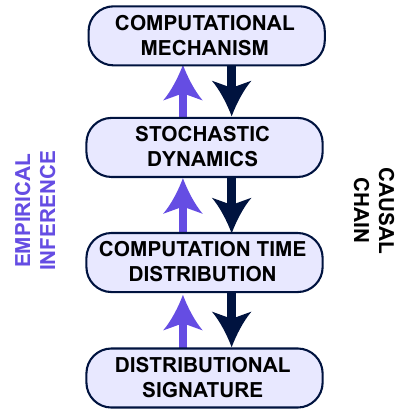}
\caption[Causal chain and empirical inference for computational mechanisms and distributional signatures]{\textbf{Causal chain and empirical inference for computational mechanisms and distributional signatures.} Computational mechanisms give rise to stochastic dynamics, which in turn generate characteristic computation time distributions and their associated distributional signatures. Empirical inference proceeds in the opposite direction, using observed distributional signatures to infer the underlying computational mechanism.}
\label{fig4} 
\end{figure}

We have shown how the stochastic dynamics of a computational system give rise to distinct computation time distributions. Because computational mechanisms causally determine stochastic dynamics, and stochastic dynamics generate characteristic computation time distributions, the reverse direction becomes an empirical inference problem. Computation time distributions and profile shapes reflect the stochastic dynamics of the system and therefore contain information about the underlying computational mechanisms. Fig. \ref{fig4} illustrates both the causal chain and the corresponding empirical inference process, in which observed computation time distributions and profile shapes are used to infer the computational mechanism that generated them.

The resulting framework provides the foundation for an empirical science of finite-scale physical computation, in which computational mechanisms are characterised through their observable behaviour and compared across diverse computational systems. Within this framework, computation time is treated as an observable rather than a computational resource. Computation time distributions are an observable consequence of the stochastic dynamics of a computational mechanism. Empirical observation therefore provides information about computational efficiency through benchmarking and mean-based analysis, and about computational mechanisms through distributional signatures. Computational complexity and the present framework are thus complementary approaches: computational complexity characterises asymptotic resources within abstract computational models, whereas the present framework investigates the computational organisation of finite physical implementations.

The framework suggests several forms of inference for the empirical study of computational systems. Temporal inference studies how computational mechanisms evolve. The computational mechanisms considered thus far have been treated as stationary, giving rise to characteristic computation time distributions and profile shapes that remain stable over time. However, some computational systems modify their computational organisation over time through processes such as learning or adaptation. \cite{lecun2015deep} In these systems, learning can be viewed as the progressive evolution of computational mechanisms, giving rise to corresponding changes in stochastic dynamics and computation time distributions. This suggests that learning may be characterised empirically through changes in distributional signatures, allowing different learning systems to be compared according to how their computational mechanisms evolve over time.

More broadly, the framework enables empirical comparison of diverse physical implementations of computation, including classical, analogue, neuromorphic, quantum, and biological systems, by comparing the computational mechanisms they employ rather than the details of their implementation. Such cross-system comparisons may reveal which computational mechanisms contribute to computational efficiency across different physical substrates.

Perhaps most importantly, the framework provides an empirical approach to the study of computational systems whose internal organisation cannot be directly observed. Rather than relying on access to internal representations or algorithms, computational mechanisms may be inferred from observable computation time distributions. This structural inference can be applied to large-scale AI systems \cite{bommasani2021opportunities}, biological cognition, and future computational systems whose internal organisation is only partially understood. Together, these forms of inference extend empirical benchmarking from measuring computational performance towards characterising and comparing computational organisation.

However, whilst there is a causal link between the stochasticity of a computational system and its distribution, we cannot claim that a distinct profile uniquely identifies the underlying mechanism. For example, it is possible that distinct mechanisms generate similar profiles. Also, datasets used to compile a profile are finite, and therefore need to be representative of an infinite dataset. Finally, it is also possible that a system uses multiple distinct computational mechanisms, resulting in an ambiguous profile. Thus, we need to apply our framework carefully.

To enable richer descriptions of computational organisation, the framework should be expanded by identifying and characterising additional computational mechanisms. This includes understanding how computational mechanisms combine and interact within hybrid computational systems, how they evolve in adaptive systems, and how similar computational processes realised in different physical implementations may nevertheless give rise to different computational mechanisms. 

By bringing together ideas from computer science, cognitive science, and the physics of computation, the framework developed here provides a common conceptual and empirical language for finite-scale physical computation. Continued development of the catalogue of computational mechanisms and their empirical characterisation will establish computational mechanisms as a common empirical level of description through which diverse physical implementations of computation can be systematically characterised and compared.

\section*{Materials and methods}
Complete implementation details are provided in the Supplementary Information.

\subsection*{Ethics} 
This research complied with all relevant ethics regulations. It has been approved by the Ethics Committee of the University of Vienna under reference number 00481. Written informed consent was obtained from all research participants before taking part in the experiments.

\subsection*{Participants}
We recruited a nonprobability sample of 94 participants (55 males, 39 females, mean age 27.4 years, age range 17-61) through the quantum physics groups and the Cognitive Science Hub at the University of Vienna, and through the Mental Calculation World Cup organisation. One participant contributed a large dataset (1398 trials) used to illustrate distributional properties and was excluded from comparative analyses. The remaining participants contributed on average 171 trials (range 150–200) for comparative analyses.

\subsection*{Visual search task}
Participants were given a two-dimensional visual task in which they had to find exactly one target, an emoticon constructed from 49 black and white pixels, within a background of black and white pixels. The input size, defined as the total number of black and white pixels, varied from 400 to nearly 320,000. The location of the target was set randomly, and each pixel in the background was randomly assigned black or white. 

Baseline measurements were taken in which the pattern was of a different colour (bright pink). This changed the task from an inefficient to an efficient search task with constant time complexity. Thus, any increase in RT with increasing input size in the baseline measurement could be attributed to both increased rendering in generating the image and larger mouse movements. RTs were then calculated as the difference between the start and end time for the task, minus the value of the baseline measurements at the set size of that task. 

\subsection*{Computational models}

\paragraph*{Drift diffusion models (DDMs).}
The DDMs were designed to complete a task similar to that of the human experiment, namely, to find a target hidden in a grid of possible locations. Log-likelihood maps that represent the possible locations are created for two conditions: target absent, the null hypothesis, and target present, the alternative hypothesis. When a sample is taken at a certain location, a random value is taken from a Gaussian distribution that is either centred on 0 if the target is absent or on 1 if the target is present. This value is then used to update the two log-likelihood maps, which in turn are used to calculate the posterior distribution, i.e. the probability of all samples at all locations given that the target is at a particular location. The model will continue the sampling and updating process until the posterior exceeds the decision threshold at any of the locations. The target is then declared present at that location. We count one model time step per sample; construction, update, and argmax selection on the global map are included in that single step. The RT measured in model time steps is taken to be equivalent to the total number of samples needed to reach the decision threshold. Our two drift diffusion models differ in sampling strategies: the random drift diffusion model chooses the location of the next sample randomly; the guided drift diffusion model chooses the location of the next sample as the location with the largest posterior. 

\paragraph*{Mechanistic neural network model (MNM).}

The MNM was implemented using the reduced instruction spiking processor or RISP \cite{plank2022case} to construct our shallow spiking neural network. We define a 90$\times$90 field of view whose input neurons fire in response to black pixels and remain inactive for white pixels. Input noise allows neurons to misfire with a small probability.

Within the field of view, target-matching filters process visual information in parallel. Each filter consists of an output neuron connected to a 7$\times$7  patch of input neurons with weights corresponding to the target pattern. When the weighted input exceeds threshold, the output neuron fires. Noise is applied to both input and output neurons, governing the probability that neurons flip from their intended binary state, either failing to fire when they should or firing when they should not. Filters are distributed according to a foveated density profile, producing a retina-like field of view.
 
In the exploration stage, the neural network scans through the image sequentially, as our human participants do by fixating locations in sequence. The location of the next placement of the field of view is chosen at random, with replacement: We do not maintain a memory of which locations have already been searched. When the field of view is placed on the image, if one or more output neurons fire, the field of view will stay in the same location and execute one more run - this is the confirmation stage. If an output neuron fires twice during these two runs, the target is considered found. This triggers the verification stage, wherein the found coordinates are verified against the target coordinates. If these match, the search is completed. If they do not match, the search will continue by placing the field of view at the next random location.  Each exploration or confirmation pass counts as one model time step, while target verification incurs a cost that scales with image size through a verification cost factor.
 
\subsection*{Relative model comparison}
We performed a relative model comparison to assess which of the three models best fits the data. We report two complementary approaches: a shape-based Kolmogorov–Smirnov (KS) analysis and a probabilistic leave-one-out (LOO) marginal log-likelihood (mLLH) analysis. For both, we first form characteristic datasets by dividing each reaction time (RT) by the number of image locations, giving RT/location. Because model outputs are in model time units and human data are in seconds, we align them with a participant-specific stretch factor \(s>0\) that rescales time.

\paragraph*{Characteristic distribution.}
Reliably estimating RT profiles at individual input sizes requires a large amount of data. To estimate RT profile shape across input sizes, each RT was divided by its corresponding input size, producing a characteristic distribution expressed as reaction time per image location. Under approximately linear RT scaling, this normalisation provides an aggregate estimate of the underlying profile shape while increasing statistical power by pooling observations across input sizes. 

\paragraph*{KS shape analysis.}
For each participant and model, we determine the stretch factor that best aligns distributional shape by minimising the two-sample KS distance between the participant’s empirical RT/location distribution (after stretching by \(s\)) and the model’s RT/location distribution. The optimisation is one-dimensional on the log-scale for numerical stability, uses bounded scalar optimisation (\(10^{-3}\) -- \(10^5\) on \(s\)), and does not require any resampling: KS natively handles unequal sample sizes. The resulting minimum KS value is recorded per participant and model. We then compute pairwise KS differences per participant using the convention `first minus second'. Because lower KS indicates better shape agreement, positive values favour the second model. We summarise paired differences across participants with means and standard errors, and report two-sided paired t-test results.

\paragraph*{Marginal log-likelihood analysis.}
To assess probabilistic fit we integrate over uncertainty in the stretch factor rather than fixing it. First, for each model we fit a Gamma distribution (positive support; location fixed at zero) to that model’s RT/location outputs to obtain its likelihood family. Next, for each participant and model we obtain a maximum a posteriori (MAP) stretch under a weakly informative Normal prior on \(\log s\); this provides stable per-participant stretch estimates without strongly constraining them. To avoid double-using the same participant’s data when scoring that participant, we construct a LOO empirical Bayes prior on \(\log s\) for each model by fitting a Normal distribution to the MAP stretches from all other participants (i.e., excluding the participant being scored). We then compute the participant-level marginal log-likelihood for each model by averaging the model’s likelihood over this LOO prior on \(\log s\). The integral is evaluated numerically using Gauss–Hermite quadrature in log-space for numerical stability. This yields one mLLH per participant and model. Pairwise model differences are formed per participant using the convention `first minus second' (so positive values favour the first model). For an overall summary we sum these differences across participants and quantify uncertainty by bootstrapping participants (10{,}000 resamples with replacement), reporting the mean, standard deviation and 95\% percentile interval of the total. 

\section*{Acknowledgments}
The authors would like to thank Borivoje Dakic, Radu Grosu, Claus Lamm, Eivinas Butkus, and JohnMark Taylor for discussions, and the Mental Calculation World Cup organization as well as the University of Vienna’s Cognitive Science Hub for their support in recruiting human research participants. This research received funding from the Austrian Science Fund (FWF) through [F7113] (BeyondC) and the Research Network TURIS. For the purpose of open access, the author has applied a CC BY public copyright licence to any Author Accepted Manuscript version arising from this submission.

\section*{Author contributions}

CvV conceived the conceptual framework, designed and performed the experiments, developed the benchmarking and distributional inference methodology, performed the analyses, and wrote the original manuscript. CS contributed the initial mechanistic neural network model. NK contributed to the development of the drift diffusion models, distributional analyses, and manuscript revisions. PW supervised the project. All authors discussed the results and commented on the manuscript.

\section*{Competing interests}

The authors declare no competing interests.

\bibliographystyle{apsrev4-2}
\bibliography{references}

\end{document}